\documentclass[12pt]{article}
\usepackage{bbold}
\usepackage{bm}
\usepackage{color}
\usepackage{dcolumn}
\usepackage{feynmf} 
\usepackage{graphics}
\usepackage{graphicx}
\usepackage{subfigure}
\usepackage{slashed}
\usepackage{placeins}

\usepackage{pdfpages}
\usepackage{eso-pic}
\usepackage{everyshi}
\usepackage{pdflscape}

\newcommand{\ep}{\varepsilon}

\def\be{\begin{equation}}
\def\ee{\end{equation}}
\def\bea{\begin{eqnarray}}
\def\eea{\end{eqnarray}}
\def\bse{\begin{subequations}}
\def\ese{\end{subequations}}
\def\bc{\begin{center}}
\def\ec{\end{center}}

\def\I{{\rm i}}

\def\D{{\rm d}}

\newcommand{\ie}{{\it i.e.}}
\newcommand{\eg}{{\it e.g.}}

\begin{document}

\begin{center}
  {\bf 
    Landau-Khalatnikov-Fradkin transformation and
    hatted $\zeta$-values
  }\\

  \vskip 0.5cm

  A.\ V.~Kotikov$^{1}$ and S.~Teber$^{2}$\\

 \vskip 0.5cm

$^1$Bogoliubov Laboratory of Theoretical Physics, Joint Institute for Nuclear Research, 141980 Dubna, Russia.\\
$^2$Sorbonne Universit\'e, CNRS, Laboratoire de Physique Th\'eorique et Hautes Energies, LPTHE, F-75005 Paris, France.
\end{center}
  

\begin{abstract}

   We show an exact formula obtained in \cite{Kotikov:2019bqo}, which relates hatted 
   and standard $\zeta$-values to all orders of perturbation theory.
   The formula is based on the  Landau-Khalatnikov-Fradkin (LKF) transformation between
the massless propagators of charged particles interacting with gauge fields,
in two different gauges.
\end{abstract}


\section{Introduction }

Consider the multi-loop structure of propagator-type functions 
(p-functions~\footnote{Following~\cite{Baikov:2018gap}, by p-functions we understand ($\overline{{\rm MS}}$-renormalized) 
Euclidean 2-point functions (that can also be obtained from 3-point functions by setting one external momentum to 
zero with the help of infra-red rearrangement) expressible in terms of massless propagator-type Feynman integrals also known as p-integrals.}). 
About three decades ago, it was noticed that all contributions proportional to $\zeta_4 = \pi^4/90$ mysteriously cancel out in the Adler 
function at three-loops~\cite{Gorishnii:1990vf}. Two decades later, it was shown that the four-loop contribution is also $\pi$-free and 
that a similar fact holds for the coefficient function of the Bjorken sum rule~\cite{Baikov:2010je}. There is by now mounting 
evidence, see, \eg, \cite{Baikov:2016tgj}-\cite{Moch:2018wjh},
that various massless Euclidean physical quantities demonstrate striking regularities in terms proportional 
to even $\zeta$-function values, $\zeta_{2n}$, {\it e.g.}, to $\pi^{2n}$ with $n$ being a positive integer.~\footnote{Notice also that, 
within a Schwinger-Dyson equation approach in fixed dimension, renormalized Euclidean massless correlators were shown to be expressed only 
in terms of odd zeta-values \cite{Kreimer:2006ua}.}
Such puzzling facts have recently given rise to the ``no-$\pi$ theorem''. The latter is based on the 
 observation~\cite{Broadhurst:1999xk,Baikov:2010hf} that the $\ep$-dependent transformation of the $\zeta$-values:
\be
\hat{\zeta}_3 \equiv \zeta_3 + \frac{3\ep}{2} \zeta_4 - \frac{5\ep^3}{2} \zeta_6,~~ \hat{\zeta}_5 \equiv \zeta_5 + \frac{5\ep}{2} \zeta_6,~~
\hat{\zeta}_7 \equiv \zeta_7\, ,
\label{hatZe}
\ee
eliminates even zetas from the expansion of four-loop p-integrals. A generalization of (\ref{hatZe}) to 5-, 6- and 7-loops is available
in Refs.~\cite{Baikov:2018wgs}-\cite{Baikov:2019zmy}.
 The results (\ref{hatZe}) and their extensions in \cite{Georgoudis:2018olj,Baikov:2019zmy}
give a possibility to predict
the terms $\sim \pi^{2n}$ in higher orders of perturbation theory (see their evaluation in
\cite{Baikov:2018wgs}-\cite{Baikov:2019zmy}).
  Note that the results \cite{Baikov:2018wgs}-\cite{Baikov:2019zmy}
  also contain multi-zeta values the consideration of which
is beyond the scope of the present study.

Remarkably, in Ref.~\cite{Kotikov:2019bqo}, an all order generalization of (\ref{hatZe}) could be achieved in a rather unexpected way: with the help of the LKF transformation~\cite{Landau:1955zz}.
The latter
 elegantly relates the QED fermion propagator in two different $\xi$-gauges (and similarly for the fermion-photon vertex). 
Its most important applications~(see \cite{Kotikov:2019bqo} and references therein)
are related to the study of the gauge covariance of QED Schwinger-Dyson 
equations and their solutions.
Other applications~\cite{Bashir:2002sp} are focused on estimating large orders of perturbation theory. 
Indeed, and this will play a crucial role in what follows, the non-perturbative nature of the LKF transformation allows to fix some of the
coefficients of the all-order expansion of the fermion propagator. Starting with a perturbative propagator in some fixed gauge, say $\eta$, all
the coefficients depending on the difference between the gauge fixing parameters of the two propagators, $\xi - \eta$, get fixed by a weak coupling expansion of 
the LKF-transformed initial one. Such estimations have been carried out for QED in various dimensions~\cite{Bashir:2002sp},
for
generalizations to brane worlds~\cite{Ahmad:2016dsb} and for more general SU(N) gauge theories~\cite{DeMeerleer:2018txc}.

Here we review the results \cite{Kotikov:2019bqo} of
usage of the LKF transformation in order to study general properties of the coefficients of the propagator. 
We
show how the transformation naturally reveals the existence of the hatted transcendental basis. Moreover, it
allows
us to extend the results of Eq.~(\ref{hatZe}) to any order in $\ep$.

\section{LKF transformation}
\label{sec:LKF:x-space}

In the following, we shall consider QED in an Euclidean space of dimension $d$ ($d=4-2\ep$).
The general forms of the fermion propagator in the momentum and $x$-space representations,  $S_F(p,\xi)$ and $S_F(x,\xi)$, in some gauge $\xi$ read:
\be
S_F(p,\xi) = \frac{1}{i\hat{p}} \, P(p,\xi) \, ,~~ S_F(x,\xi) =  \hat{x} \, X(x,\xi) \, ,
\label{SFp}
\ee
where the tensorial structure, \eg, the factors $\hat{p}$ and $\hat{x}$ containing Dirac $\gamma$-matrices, have been extracted.
The two representations, $S_F(x,\xi)$ and $S_F(p,\xi)$, are related by the Fourier transform which is defined as:
\be
S_F(p,\xi) = \int \frac{\D^dx}{(2\pi)^{d/2} } \, e^{\I px} \, S_F(x,\xi) \, , ~~
S_F(x,\xi) = \int \frac{\D^dp}{(2\pi)^{d/2} } \, e^{-\I px} \, S_F(p,\xi) \, .
\label{SFp2x}
\ee

The famous LKF transformation connects in a very simple way the fermion propagator in two different gauges, \eg, $\xi$ and $\eta$. In dimensional regularization, it reads \cite{Kotikov:2019bqo}:
\be
S_F(x,\xi) = S_F(x,\eta)\, e^{\I D(x)} \, .
\label{LKFN}
\ee

We may now proceed in calculating $D(x)$. In order to do so, it is possible to use the following simple formulas
for the Fourier transform of massless propagators (see, \eg,
\cite{Kotikov:2018wxe}):
\be
\int \D^dx \, \frac{e^{\I px}}{x^{2\alpha}} \, = \frac{2^{2\tilde{\alpha}} \pi^{d/2} a(\alpha)}{p^{2\tilde{\alpha}}},~~
\int \D^dp \, \frac{e^{-\I px}}{p^{2\alpha}} \, = \frac{2^{2\tilde{\alpha}} \pi^{d/2} a(\alpha)}{x^{2\tilde{\alpha}}}, ~~
a(\alpha)=\frac{\Gamma(\tilde{\alpha})}{\Gamma(\alpha)},~~\tilde{\alpha}=\frac{d}{2}-\alpha \, ,
\label{SFx2pN}
\label{SFp2xN}
\ee
This yields
with the parameter $\ep$ made explicit:
\be
D(x)=\frac{\I\,\Delta\,A}{\ep}\,\Gamma(1-\ep)\, (\pi\mu^2 x^2)^{\ep} , ~~ A=\frac{\alpha_{\rm em}}{4\pi}=\frac{e^2}{(4\pi)^2}\, .
\label{DxN}
\ee
From Eq.~(\ref{DxN}), we see that $D(x)$ contributes with a common factor $\Delta A$ accompanied by the singularity $\ep^{-1}$.

\section{LKF transformation in momentum space}
\label{sec:LKF:p-space:F}

Let's assume that, for some gauge fixing parameter $\eta$,
the fermion propagator $S_F(p,\eta)$ with external momentum $p$ has the form (\ref{SFp}) with $P(p,\eta)$ reading:
\be
P(p,\eta) = \sum_{m=0}^{\infty} a_m(\eta)\, A^m \,{\left(\frac{\tilde{\mu}^2}{p^2}\right)}^{m\ep} \, ,~~ \tilde{\mu}^2= 4\pi \mu^2 \, ,
\label{Peta}
\ee
where
$a_m(\eta)$ are coefficients of the loop expansion of the propagator and
$\tilde{\mu}$ is the renormalization scale,
which lies somehow between the MS-scale $\mu$ and the $\overline{{\rm MS}}$-scale $\overline{\mu}$.
Then, the LKF transformation shows that, for another gauge parameter $\xi$,
the fermion propagator can be expressed as:
\be
\hspace{-5pt}P(p,\xi) = \sum_{m=0}^{\infty} a_m(\xi)\, A^m\, {\left(\frac{\tilde{\mu}^2}{p^2}\right)}^{m\ep} \, ,
\label{Pxi} 
\ee
where
\be
a_m(\xi) = a_m(\eta) \, \frac{\Gamma(2-(m+1)\ep)}{\Gamma(1+m\ep)}\,
\sum_{l=0}^{\infty} 
\frac{\Gamma(1+(m+l)\ep)\,\Gamma^l(1-\ep)}{l!\,\Gamma(2-(m+l+1)\ep)} \, \frac{(\Delta \, A)^l}{(-\ep)^l} \, {\left(\frac{\tilde{\mu}^2}{p^2}\right)}^{l\ep} \, .
\label{axi}
\ee
In order to derive (\ref{axi}), we used the fermion propagator $S_F(p,\eta)$ with  $P(p,\eta)$ given by (\ref{Peta}), did the Fourier transform to
$S_F(x,\eta)$ and applied the LKF transformation (\ref{LKFN}). As a final step, we took the inverse Fourier transform and obtained 
$S_F(p,\xi)$ with  $P(p,\xi)$ given by (\ref{Pxi}).

\subsection{Scale fixing}

Following  \cite{Kotikov:2019bqo},
we consider only the case of the so-called MS-like schemes.
In such schemes, we need to fix specific terms coming from the application of dimensional regularization. Such a procedure will be called
{\it scale fixing} 
and will play a crucial role in our analysis.
	
Let's first recall that the $\overline{{\rm MS}}$-scale $\overline{\mu}$ is related to the previously defined scale $\tilde{\mu}$ with the help of
$\overline{\mu}^2=\tilde{\mu}^2\, e^{-\gamma}$,
where $\gamma$ is the Euler constant. An advantage of the $\overline{{\rm MS}}$-scale is that it subtracts the Euler constant $\gamma$ from the $\ep$-expansion.
Moreover, it is well known that, in calculations of two-point massless diagrams, the final results do not display any $\zeta_2$.
~\footnote{Strictly speaking, $\zeta_2$ can appear in some formulas such as sum rules in deep-inelastic scattering. They originate from 
an analytic continuation \cite{Kotikov:2005gr} of certain special forms of p-integrals. We will not consider this case in the present study.}
 So it is convenient to choose some scale which also subtracts $\zeta_2$ in intermediate steps of the calculation.
 For this purpose, in \cite{Kotikov:2019bqo} we
 considered two different scales. 
	
 The first one is the popular $G$-scale \cite{Chetyrkin:1980pr}.
Actually, following \cite{Broadhurst:1999xk}, in Ref.~\cite{Kotikov:2019bqo}
we
used a slight modification of this scale that we
refer to as the $g$-scale
and in which an additional factor $1/(1-2\ep)$ is subtracted from the one-loop result.

Moreover, in
\cite{Kotikov:2019bqo} we
also introduced a new scale which is based on old calculations of massless diagrams performed by Vladimirov who added \cite{Vladimirov:1979zm}
an additional factor $\Gamma(1-\ep)$ to each loop contribution. The latter corresponds to adding the factor $\Gamma^{-1}(1-\ep)$ to the corresponding scale. 
We shall refer to this scale as the minimal Vladimirov-scale, or MV-scale, and define:~\footnote{Notice that the form (\ref{Vla}) has been used once to define the
$\overline{{\rm MS}}$ scheme (see Errata to Ref.~\cite{Kataev:1988sq}).}
\be
\mu_{{\rm MV}}^{2\ep}=\frac{\tilde{\mu}^{2\ep}}{\Gamma(1-\ep)} \, .
\label{Vla}
\ee
The use of the MV-scale leads to simpler results in comparison with the $g$ one. Hence, the MV-scale is more appropriate to our analysis and 
all our
results are
given in the MV-scale.
Differences coming from the use of the $g$-scale can be found in Ref. \cite{Kotikov:2019bqo}.

In
the MV-scale,
we can rewrite the result (\ref{axi}) in the following general form:~\footnote{The results in the case of scalar QED are very similar and can be found in Ref. \cite{Kotikov:2019bqo}.}
\be
a_m(\xi) = a_m(\eta)
\sum_{l=0}^{\infty} \, \frac{1-(m+1)\ep}{1-(m+l+1)\ep} \,
\Phi_{\rm MV}(m,l,\ep) 
\, \frac{(\Delta \, A)^l}{(-\ep)^l l!} \, {\left(\frac{\mu_{\rm MV}^2}{p^2}\right)}^{l\ep} \, ,
\label{axi.1}
\ee
where
\be
\Phi_{{\rm MV}}(m,l,\ep)=\frac{\Gamma(1-(m+1)\ep)\Gamma(1+(m+l)\ep)\Gamma^{2l}(1-\ep)}{
 \Gamma(1+m\ep)\Gamma(1-(m+l+1)\ep)} \, .
\label{Phi:V:def}
\ee
In Eq.~(\ref{axi.1}), the factor $(1-(m+1)\ep)/(1-(m+l+1)\ep)$ has been specially extracted from $\Phi_{\rm MV}(m,l,\ep)$ in order to insure 
equal transcendental level, \ie, the same value of $s$ for $\zeta_s$ at every order of the $\ep$-expansion of $\Phi_{\rm MV}(m,l,\ep)$ (see below).

\subsection{MV-scale}

The $\Gamma$-function $\Gamma(1+\beta\ep)$ has the following expansion:
\be
\Gamma(1+\beta\ep) = \exp \Big[ -\gamma \beta \ep + \sum_{s=2}^{\infty}\, (-1)^s \, \eta_s \beta^s \ep^s \Bigr],~~ 
\eta_s = \frac{\zeta_s}{s} \, .
\label{Gamma:exp}
\ee
Substituting Eq.~(\ref{Gamma:exp}) in Eq.~(\ref{Phi:V:def}), yields for the factor $\Phi_{{\rm MV}}(m,l,\ep)$:
\be
\Phi_{{\rm MV}}(m,l,\ep)= \exp \Big[ \sum_{s=2}^{\infty}\,\eta_s \, p_s(m,l) \, \ep^s \Bigr]\, ,
\label{Phi:V}
\ee
where
\be
p_s(m,l)=  (m+1)^s-(m+l+1)^s + 2l + 
(-1)^s \Bigl\{(m+l)^s-m^s\Bigr\}\, ,
\label{ps:V}
\ee
and, as expected from the MV-scale, we do have:
\be
p_1(m,l)=0,\qquad p_2(m,l)=0 \, .
\label{ps:V:1-2}
\ee

As can be see from Eq.~(\ref{Phi:V}), $\Phi_{{\rm MV}}(m,l,\ep)$ contains $\zeta_s$-function values of a given weight (or transcendental level) $s$ in factor of $\ep^s$. 
Such a property strongly constrains the coefficients of the $\ep$-series thereby simplifying our analysis. It is reminiscent of the one earlier found in Ref.~\cite{Kotikov:2000pm}.
When judiciously used, it sometimes allows to derive results without any calculations (as in Ref.~\cite{Kotikov:2002ab}).
In other cases, it simplifies the structure of the results which can then be predicted as an ansatz in a very simple way
(see Refs.~\cite{Fleischer:1998nb,Kotikov:2007cy}). For a recent application of such property, see the recent papers \cite{Dixon:2019uzg} and references and discussions therein.

\section{Solution of the recurrence relations}
\label{sec:LKF:proof}

We now focus on the polynomial $p_s(m,l)$ of Eq.~(\ref{ps:V}) that is conveniently separated in even and odd $s$ values. Then, we see that the following recursion relations hold:
\be
	p_{2k} = p_{2k-1} + L p_{2k-2} + p_{3}, ~~ p_{2k-1} = p_{2k-2} + L p_{2k-3} + p_{3}, ~~  L=l(l+1) \, . 
\label{p2k-1:V}
\ee
Specific to the MV-scheme, these relations only depend on $L$ which leads to strong simplifications. Nevertheless, they are difficult to solve for arbitrary $k$. It is simpler 
to proceed by explicitly considering the first values of $k$:
\be
p_4= 2p_3\, , ~~ 
p_5= p_4 + Lp_3 + p_3 = (3+L)p_3\, , ~~
p_6= p_5 + Lp_4 + p_3 = (4+3L)p_3\, ,
\label{p6:V}
\ee
showing that $p_s$ takes the form of a polynomial in $L$ in factor of $p_3$.
Then, taking the results 
in (\ref{p6:V}) together, yields:
\be
Lp_3=p_5-3p_3, \qquad p_6=3p_5-5p_3 \, ,
\label{p6:V:express}
\ee
which reveals that the even polynomial $p_6$ can be entirely expressed in terms of the lower order odd ones, $p_3$ and $p_5$.
We may automate this procedure for higher values of $k$ and express $p_{2k}$ as
\be
p_{2k}=\sum_{s=2}^{k} p_{2s-1} \, C_{2k,2s-1} \, =  \sum_{m=1}^{k-1} p_{2k-2m+1} \, C_{2k,2k-2m+1} \, .
\label{p2k:V:general}
\ee
From these results, it is possible to determine the exact $k$-dependence
of $C_{2k,2s-1}$, which has the following structure:
\be
C_{2k,2k-2m+1} = b_{2m-1}
\, \frac{(2k)!}{(2m-1)! \, (2k-2m+1)!} \, ,
\label{C:V:structure}
\ee
with the first coefficients $b_{2m-1}$ taking the values:
\bea
&&\hspace{-1cm} b_{1}= \frac{1}{2},~ b_{3}= -\frac{1}{4},~ b_{5}= \frac{1}{2},~ b_{7}= -\frac{17}{2},~
b_{9}= \frac{31}{2},~
b_{11}= -\frac{691}{4},~ b_{13}= \frac{5461}{2},~
b_{15}= -\frac{929569}{16},
\nonumber \\
&&\hspace{-1cm} b_{17}= \frac{3202291}{2},~ b_{19}= -\frac{221930581}{4},~
b_{21}= \frac{4722116521}{2},~ b_{23}= -\frac{968383680827}{8} \, .
\label{b:V:values}
\eea
Examining the numerators of $b_{2m-1}$, one can see that they are proportional to the numerators of Bernoulli
numbers. Indeed, a closer inspection reveals that, accurate to a sign, the coefficients $b_{2m-1}$
coincide with the zero values of Euler polynomials $E_n(x)$:
\be
b_{2m-1} \, = - E_{2m-1}(x=0)\, ,
\label{b:V:expressionE}
\ee
and therefore to Bernoulli and Genocchi numbers, $B_m$ and $G_m$, respectively, because
\be
E_{2m-1}(x=0) = \frac{G_{2m}}{2m}, \quad G_{2m} = - \frac{(2^{2m} - 1)}{m} \, B_{2m}\, .
\label{Ex=0}
\ee
Hence, the compact formula for the coefficients $b_{2m-1}$, expressed through the well known Bernoulli numbers $B_m$, reads:
\be
b_{2m-1} = \frac{(2^{2m} - 1)}{m} \, B_{2m} \, .
\label{b:V:expressionB}
\ee
Together with (\ref{C:V:structure}), Eq.~(\ref{b:V:expressionB}) provides an exact analytic expression for $p_{2k}$, Eq.~(\ref{p2k:V:general}), 
for arbitrary values of $k$.

\section{Hatted $\zeta$-values}


At this point, it is convenient to represent the argument of the exponential in the r.h.s.\ of (\ref{Phi:V}) as follows:
\be
\sum_{s=3}^{\infty}\,\eta_s \, p_s \, \ep^s = \sum_{k=2}^{\infty}\,\eta_{2k} \, p_{2k} \, \ep^{2k} +
\sum_{k=2}^{\infty}\,\eta_{2k-1} \, p_{2k-1} \, \ep^{2k-1} \, .
\label{Phi:V:exp}
\ee
With the help of Eq.~(\ref{p2k:V:general}), the first term in the r.h.s.\ of Eq.~(\ref{Phi:V:exp}) may be expressed as:
\be
\sum_{k=2}^{\infty}\,\eta_{2k} \, p_{2k} \, \ep^{2k} = \sum_{k=2}^{\infty}\,\eta_{2k}  \, \ep^{2k} \, \sum_{s=2}^k p_{2s-1} \,
C_{2k,2s-1} =
\sum_{s=2}^{\infty} p_{2s-1} \, \sum_{k=s}^{\infty}\,\eta_{2k} \, C_{2k,2s-1} \, \ep^{2k} \, .
\label{Phi:V:exp1}
\ee
Then, Eq.~(\ref{Phi:V:exp}) can be written as $\sum_{s=2}^{\infty}\,\hat{\eta}_{2s-1} \, p_{2s-1} \, \ep^{2s-1}$
where
\be
\hat{\eta}_{2s-1} = \eta_{2s-1} +  \sum_{k=s}^{\infty}\,\eta_{2k} \, C_{2k,2s-1} \, \ep^{2(k-s)+1},~~
C_{2k,2s-1} = b_{2k-2s+1} \, \frac{(2k)!}{(2s-1)! \, (2k-2s+1)!} \, .
\label{hEta}
\ee
Thus, Eq.~(\ref{Phi:V}) can be represented as:
\be
\Phi_{{\rm MV}}(m,l,\ep) = \exp \Big[ \sum_{s=2}^{\infty}\,\hat{\eta}_{2s-1} \, p_{2s-1} \, \ep^{2s-1} \Bigr]
= \exp \Big[ \sum_{s=2}^{\infty}\,\frac{\hat{\zeta}_{2s-1}}{2s-1} \, p_{2s-1} \, \ep^{2s-1} \Bigr] \, ,
\label{Phi:V2}
\ee
where
\be
\hat{\zeta}_{2s-1} = \zeta_{2s-1} +  \sum_{k=s}^{\infty}\,\zeta_{2k} \, \hat{C}_{2k,2s-1} \, \ep^{2(k-s)+1} 
\label{hZeta}
\ee
with
\be
\hat{C}_{2k,2s-1} = \frac{2s-1}{2k} \, C_{2k,2s-1}  
= b_{2k-2s+1} \, \frac{(2k-1)!}{(2s-2)! \, (2k-2s+1)!} 
\, .
\label{hC}
\ee
Together with (\ref{hC}) and (\ref{b:V:expressionB}), Eq.~(\ref{hZeta}) provides an exact expression for the hatted $\zeta$-values in terms of the standard
ones valid for all $\ep$.

\section{Summary}
\label{sec:summary}

From the result (\ref{axi.1}) corresponding to the LKF transformation of the fermion propagator
we have found peculiar recursion relations (\ref{p2k-1:V}) between even and odd values of the polynomial associated to the 
uniformly transcendental factor $\Phi_{{\rm MV}}(m,l,\ep)$ (\ref{Phi:V:def}). These relations are simple in the MV-scheme
that we have introduced in Eq. (\ref{Vla}).
They relate the even and odd parts in a rather simple way (see
(\ref{p2k:V:general})) which reveals the possibility (\ref{Phi:V2})
to express all results for $\Phi_{{\rm MV}}(m,l,\ep)$ in terms of hatted $\zeta$-values.
Our careful study of the recursion relations (\ref{p2k-1:V})
allowed us to derive exact formulas, Eqs.~(\ref{hEta}) and (\ref{hZeta}), relating hatted and standard $\zeta$-values to all orders of perturbation theory. 
The coefficients of the relations are expressed trough the well-known Bernoulli numbers, $B_{2m}$
(see
(\ref{hC}) and (\ref{b:V:expressionB})).
Our results provide
stringent constraints on multi-loop calculations at any order in perturbation theory.\\


One of us (A.V.K.)
thanks the Organizing Committee of
International Bogolyubov Conference
``Problems of Theoretical and Mathematical Physics''
 for their invitation.

\end{document}